\documentstyle[prc,aps,epsfig,twocolumn,eqsecnum]{revtex}

\catcode`\@=11

\def\maketitle2{\par 
\begingroup
\let\cite\@bylinecite
\def\thefootnote{\fnsymbol{footnote}}%
\twocolumn[\@maketitle2\vskip2pc]%
\thispagestyle{plain}\@thanks
\endgroup
\def\thefootnote{\arabic{footnote}}%
\setcounter{footnote}{0}%
\let\maketitle2\relax \let\@maketitle2\relax
\let\@thanks\relax \let\@authoraddress\relax \let\@title\relax
\let\@date\relax \let\thanks\relax \let\@abstract\relax 
\let\@pacs\relax}

\def\abstract#1{\gdef\@abstract{{\par 
\bgroup
\ifdim\prevdepth=-1000pt \prevdepth0pt\fi
\hsize\columnwidth
\dimen0=-\prevdepth \advance\dimen0 by17.5pt \nointerlineskip
\small\vrule width 0pt height\dimen0 \relax}{~~}#1\egroup}}

\def\pacs#1{\gdef\@pacs{{\par 
\bgroup
\hsize\columnwidth \parindent0pt
\ifdim\prevdepth=-1000pt \prevdepth0pt\fi
\dimen0=-\prevdepth \advance\dimen0 by20pt\nointerlineskip
\egroup} PACS numbers:~#1}}

\def\@maketitle2{
\@preprint
\@title
\ifdim\prevdepth=-1000pt \prevdepth0pt\fi
\@authoraddress
\@date
\begin{list}{}{\leftmargin=0.10753\textwidth \rightmargin=\leftmargin
\itemsep=1pc\partopsep=-1pc}
\item\@abstract
\item\@pacs
\end{list}
}

\catcode`\@=12

\newcommand{\nph}[1]{#1$p$#1$h$}

\begin{document}
%
%
\title{Ground state correlations and mean-field in $^{16}$O: \\
  Part II} 

\author{Bogdan~Mihaila(a,b)
        \thanks{electronic mail:Bogdan.Mihaila@unh.edu} and
        Jochen~H.~Heisenberg(a)
        \thanks{electronic mail:Jochen.Heisenberg@unh.edu} }

\address{\ \\
   (a) Department of Physics, University of New Hampshire,
       Durham, NH 03824 \\
   (b) Chemistry and Physics Department,
       Coastal Carolina University, Conway, SC 29526
        }
\date{\today}

\abstract{ We continue the investigations of the $^{16}$O ground state
  using the coupled-cluster expansion [$\exp({\bf S})$] method with
  realistic nuclear interaction.  In this stage of the project, we
  take into account the three nucleon interaction, and examine in some
  detail the definition of the internal Hamiltonian, thus trying to
  correct for the center-of-mass motion.  We show that this may result
  in a better separation of the internal and center-of-mass degrees of
  freedom in the many-body nuclear wave function.  The resulting
  ground state wave function is used to calculate the ``theoretical"
  charge form factor and charge density.  Using the ``theoretical"
  charge density, we generate the charge form factor in the DWBA
  picture, which is then compared with the available experimental
  data.  The longitudinal response function in inclusive electron
  scattering for $^{16}$O is also computed.  }

\maketitle2 


\section{Introduction}

The coupled-cluster expansion, also called the $\exp({\bf S})$ method,
was developed about 40 years ago by Coester~\cite{ref:Coester}. It was
then applied to finite nuclei quite successfully by K\"ummel and the
Bochum group~\cite{ref:Kummel_etal} using a representation of the wave
function in coordinate space together with common interactions of that
time. While the method in general is viewed as essentially exact,
approximations are introduced stemming from truncations in the coupled
cluster equations, as well as truncations in the model space.  The
results for the binding energy were well above the Coester-line and
were taken as evidence for the presence of three-nucleon and higher
order interactions.

While few further developments took place since that time, the ongoing
expansion in computer power as well as the availability of more
sophisticated nucleon-nucleon interactions suggests that one should go
back to these precise methods.  During the past several years we have
reexamined the coupled cluster approach and applied it to the
spherical nucleus $^{16}$O using the Argonne $v$18
potential~\cite{ref:paper_one}. We combine the mean field approach
together with the coupled cluster expansion in a harmonic oscillator
basis of 50 $\hbar \omega$ (which constitutes our truncated Hilbert
space) and going up to the level of \nph{4} clusters.

Further, we have expanded the formulation to include the Urbana IX
three-nucleon interaction. The results are in reasonable agreement
with the experiment and make the subject of the present paper.  These
results obtained using the coupled cluster expansion should be
directly comparable with those obtained using the Variational Monte
Carlo (VMC)~\cite{ref:Pieper_etal} or the Green's Function Monte Carlo
(GFMC) methods, using the same interaction. Especially the GFMC method
with the Argonne $v$18 potential and the Urbana IX three-nucleon
interaction has proved quite accurate in nuclei with $A\le
8$~\cite{ref:Pudliner_etal}, even though a certain uncertainty in the
three-nucleon interaction model is responsible for a systematic lack
of binding in the heavier systems. These difficulties will be
addressed however in the future, by considering more sophisticated
three-body interaction models.

In section~\ref{sec:CCM} we briefly review the coupled-cluster
approach as outlined in~\cite{ref:paper_one} when the Hamiltonian
includes only the two-body interaction. We continue this discussion in
section~\ref{sec:HCM} with comments pertaining to the inclusion of the
center-of-mass Hamiltonian, which is supposed to constrain the
center-of-mass degrees of motion in the ground state wave function. In
section~\ref{sec:3b} we outline the changes necessary in order to
include the three-nucleon interaction via a density-dependent
approximation. In section~\ref{sec:obs} we show various observables
calculated in the ground state for $^{16}$O. Finally, in
section~\ref{sec:end} we present our conclusions and a brief outlook.

%
%
\section{Coupled-cluster method: a review}
\label{sec:CCM}

For a spherically symmetric nuclear system, consisting of both 
protons and neutrons, the total Hamiltonian is given as
the sum of a nonrelativistic one-body kinetic energy,
a two-nucleon potential and a three-nucleon potential 
\begin{eqnarray}
   {\bf H} 
   & = & 
   \sum_i T_i 
   \ + \ \sum_{i<j} \ V_{ij} 
   \ + \ \sum_{i<j<k} \ V^{tni}_{ijk}
   \>.
\label{eq:Hamiltonian_0}
\end{eqnarray}
Since our calculation is carried out in configuration space, we need
the nuclear interaction projected out in an operator format allowing
for calculation of two- and three-body matrix elements.  We chose to
use the operator format of the Argonne and Urbana family potentials.
The Argonne $v$18 model~\cite{ref:argonne_v18} is one of a new class
of $NN$ potentials that accurately fit both $pp$ and $nn$ scattering
data up to 350 MeV with a $\chi^2$/datum near one. The Urbana IX
potential~\cite{ref:tnipot} includes a long-range two-pion exchange
and a short-range phenomenological component. The strength
constant of the short-range component is adjusted to reproduce the
binding energy of the three-nucleon system.

In configuration space, the many-body correlated ground state,
$|\tilde 0 \rangle$, of $H$ is written as a linear combination of all
possible $n$ particle configurations, which are defined in terms of
the \nph{n} creation operators, (${\bf O}^{\dag}_0$~=~${\bf 1}$, ${\bf
  O}^{\dag}_1$~=~${\bf a}^{\dag}_{p_1} {\bf a}_{h_1}$, ${\bf
  O}^{\dag}_2$~=~${\bf a}^{\dag}_{p_1} {\bf a}^{\dag}_{p_2} {\bf
  a}_{h_2} {\bf a}_{h_1}$) acting on the reference state, $|0\rangle$.
Since the nucleon system obeys the Fermi statistics, the natural
choice for the vacuum $|0\rangle$ is a $n$ single-particle Slater
determinant.  In the coupled-cluster expansion, we have
\[
   | \tilde{0} \rangle \ = \ {\displaystyle e^{{\bf S}^{\dag}}} | 0 \rangle
   \>,
\]
where ${\bf S}^{\dag}$ is the cluster correlation operator, defined in
terms of its {\em ph}-creation operators expansion as
\[
   {\bf S}^{\dag} = \sum_{n=0}^\infty \frac{1}{n!} S_n {\bf O}^{\dag}_n
   \>.
\]
In order to determine the amplitudes $S_n$ we solve a set of
non-linear equations, which may be obtained 
using a variational principle~\cite{ref:paper_one}:
\begin{eqnarray}
   \langle 0 | \ e^{\bf S} \ {\bf H} \ e^{- \, {\bf S}} \ 
   {\bf O}^{\dag}_n \
   | 0 \rangle
   & = & 0
   \>,
   \qquad 
   n \ge 1
   \>.
\label{eq:ccm_eqns}
\end{eqnarray}
The coefficients of the operators
$\{ {\bf S}_n \}$ carry the physical significance of 
nuclear correlations.

In reference~\cite{ref:paper_one} we have outlined our approach to
solving the coupled-cluster equations~(\ref{eq:ccm_eqns}).  We first
introduce a mean-field Hamiltonian in terms of the one-body kinetic
energy operator and a mean-field potential. The mean-field is
partially constrained by the assumption that ${\bf S}_1 = 0$, or the {\em
  maximum overlap basis}.  The set of single-particle energies and
wave functions which determine the configuration space is determined by
diagonalizing the mean-field Hamiltonian.  

The mean-field component calculation is carried out together with the
calculation of the \nph{2} correlations in a self-consistent manner.
Even though the \nph{3} and \nph{4} correlations are not calculated
explicitly, the \nph{2} correlations are implicitly corrected for the
presence of higher-order correlations, consistently with our
truncation scheme. In this respect we replace the traditional
truncation in ${\bf S}_n$~\cite{ref:Kummel_etal} by a truncation
in $1/\epsilon_n$, where $\epsilon_n$ is the \nph{n} excitation
energy. The results presented here are obtained by keeping only terms
up to second order in $1/\epsilon$.

%
%
\section{Center-of-mass Hamiltonian}
\label{sec:HCM}

The correlated ground-state $| \tilde 0 \rangle$ is a function of $A$
sets of coordinates $\{\vec r_k, k = 1\ldots A\}$. Consequently, the
ground state wave function is not translationally invariant.  As a
result, it is common practice~\cite{ref:paper_one} to attempt solving
the Schr\"odinger equation for the ground state of the {\em internal}
Hamiltonian
\[
   H_{int} 
   \ = \ H \ - \ T_{CM}
   \>,
\]
instead of the Hamiltonian~(\ref{eq:Hamiltonian_0}).  The {\em
  internal} Hamiltonian is entirely written in the center-of-mass
frame by removing the center-of-mass kinetic energy, $T_{CM} =
P_{CM}^2/(2 \, m \, A)$, with $m$ being the nucleon mass.  Both the
two- and three-nucleon interactions are given in terms of the relative
distances between nucleons, so in this respect no corrections are
needed.

However, this does not prevent the contamination of the correlated
ground-state $| 0 \rangle$ by the center-of-mass motion.  Especially
since we are interested in the calculation of the excited states
spectrum, this contamination will result in spurious excitations of
the center-of-mass. Moreover, the calculation of observables to be
compared with experimental data which are taken in the center-of-mass
frame, requires the evaluation of a many-body expansion as explained
in~\cite{ref:paper_two}. This in turn relies on the assumption that we
can neglect the correlations between the internal and center-of-mass
degrees of freedom at the level of the correlated ground-state $| 0
\rangle$.
In order to ensure such a separation, a supplemental center-of-mass
Hamiltonian is added
\[
  H_{CM} \ = \ 
  \beta_{CM} \, 
   \left [ T_{CM} \ + \ \frac{1}{2} \ (m \, A) \ \Omega^2 \ R_{CM}^2
   \right ]
  \>.
\]   
The center-of-mass Hamiltonian has the role of constraining the
center-of-mass component of the ground state wave
function~\cite{ref:CM_people}.  Then, we have to identify a domain of
values for the parameters $\beta_{CM}$ and $\Omega^2$ such that the
binding energy
\[
E \ = \ \langle H_{int}^{'} = H_{int} + H_{CM} \rangle 
            \, - \, \langle H_{CM} \rangle
\]             
is insensitive to the choice of these parameters.

\begin{figure}[t]
   \epsfxsize = 3.0in
   \centerline{\epsfbox{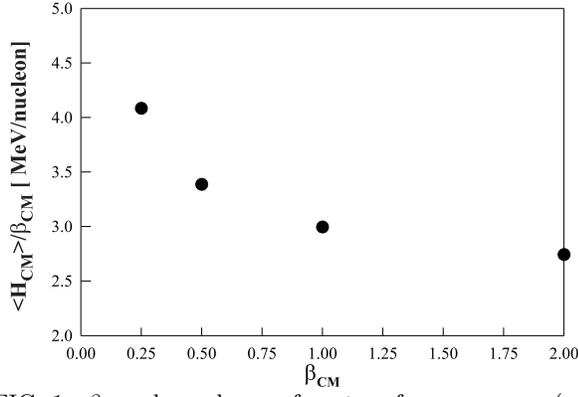}}
   \caption{$\beta_{CM}$ dependence of center-of-mass energy (using $v$18).}
\label{fig:vcm_v18}
\end{figure}

\begin{figure}[b]
   \epsfxsize = 3.0in
   \centerline{\epsfbox{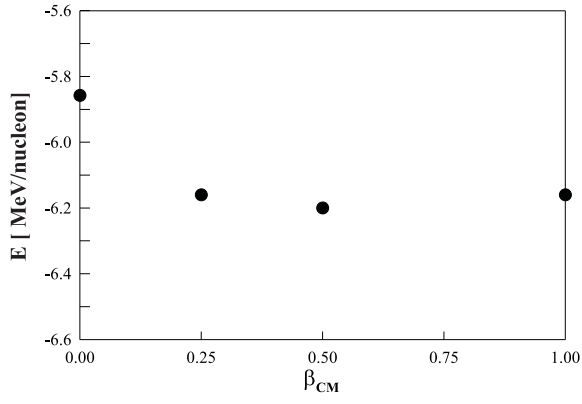}}
   \caption{$\beta_{CM}$ dependence of binding energy (using $v$18).}
\label{fig:evcm_v18}
\end{figure}

For the purpose of this analysis we leave out the three-nucleon
interaction, and show results when including only the $v$18 $NN$
interaction.  Figures~\ref{fig:vcm_v18} and~\ref{fig:evcm_v18} show
the $\beta_{CM}$ dependence of center-of-mass and binding energies,
respectively.

%
%
\section{Three-nucleon interaction corrections}
\label{sec:3b}

We will present here the corrections necessary to take into account
the three-nucleon interaction as part of the general nuclear
interaction.  As discussed before we consider the three-nucleon
interaction as the sum of two components: a long-range two-pion
exchange and a short-range phenomenological component. The calculation
of the three-nucleon interaction has been presented
elsewhere~\cite{ref:3b-me}.

Given the form~(\ref{eq:Hamiltonian_0}) of the Hamiltonian, we write
the operator $V_{tni}$ in second quantization as
\begin{equation}
   {\bf V}_{tni}
   = 
   \sum_{a_1b_1c_1a_2b_2c_2} 
       V_{a_1b_1c_1,a_2b_2c_2} 
       {\bf a}^{\dag}_{a_1}
       {\bf a}^{\dag}_{b_1}
       {\bf a}^{\dag}_{c_1}
       {\bf a}_{c_2}
       {\bf a}_{b_2}
       {\bf a}_{a_2}
   \>.
\end{equation}
Here, the matrix elements are given as integrals involving the single
particle states (including spins)
\begin{eqnarray}
   &&
   V_{a_1b_1c_1,a_2b_2c_2} 
   \nonumber \\ &&
   = \
   \langle \phi_{a_1}(1) \phi_{b_1}(2) \phi_{c_1}(3)
       | V_{tni} | 
           \phi_{a_2}(1) \phi_{b_2}(2) \phi_{c_2}(3) \rangle 
   \>.
\label{eq:vtni_comp}
\end{eqnarray}
In addition to being symmetric with respect to the interchange of the
particles labelled $2$ and $3$
\begin{equation}
   V_{a_1b_1c_1,a_2b_2c_2}
   \ = \
   V_{a_1c_1b_1,a_2c_2b_2}
   \>,
\label{eq:vtni_symm_a}
\end{equation}
the integrals~(\ref{eq:vtni_comp}) also satisfy the symmetry
\begin{equation}
   V_{a_1b_1c_1,a_2b_2c_2}
   \ = \
   V_{c_1a_1b_1,c_2a_2b_2}
   \>.
\label{eq:vtni_symm_b}
\end{equation}
The last property is a consequence of the cyclic sums involved in the
definition of the interaction, which make the interaction invariant
with respect to the labelling of the particles.

In previous applications the approximation has been made that such an
interaction can be represented by a density dependent two-body
interaction.  While such a substitution is the easiest modification,
it has been stated that this is insufficient~\cite{ref:tnipot}.
However, the rigorous inclusion of the three-nucleon interaction in
configuration space using the coupled-cluster method is seriously
hampered by our present computing capabilities and the necessary size
of the configuration space.  Ideally, we would like to calculate all
integrals of the form (\ref{eq:vtni_comp}) without any artificial
restrictions.  In practice though, we must limit ourselves to
calculating matrix elements of the form
\begin{equation}
   V_{\alpha \, a_1b_1, \beta \, a_2b_2} 
   \quad \rm{and} \quad 
   V_{\alpha \, a_1b_1, a_2 \, \beta \, b_2}
   \quad \>,
\label{eq:vtni_calc}
\end{equation} 
where $\alpha$ and $\beta$ couple to spin 0 only.
We are then faced with a compromise:
Since matrix elements of the form (\ref{eq:vtni_calc}) are all we need
in order to calculate exactly the first- and second-order (${\bf
  S}_2$) contributions to the mean-field and binding energy, the
leading orders in our expansion are treated rigorously correct.  Then
we make a reduction of the three-nucleon interaction to an effective
two-body interaction, and use this effective interaction when dealing
with higher-order corrections (${\bf S}_n, \, n\ge$3).  This is
achieved by defining the effective two-body interaction as
\begin{eqnarray}
   {\bf V}_{tni, den} \ = \ 
   \frac{1}{4} \, 
   \sum_{a_1b_1a_2b_2} \, 
       V^{tni, den}_{a_1b_1,a_2b_2} \,  
       {\bf a}^{\dag}_{a_1}
       {\bf a}^{\dag}_{b_1}
       {\bf a}_{b_2}
       {\bf a}_{a_2}
   \>,
\end{eqnarray}
where we define the matrix element $V^{tni, den}_{a_1b_1,a_2b_2}$ to
be equal to
\begin{eqnarray}
   &&
   \langle 0 | 
      {\bf V}^{tni}  
      {\bf a}_{a_1}^{\dag} {\bf a}_{a_2}^{\dag} {\bf a}_{b_2} {\bf a}_{b_1}
      | 0 \rangle
   \nonumber \\ && 
   =
      \sum_h \biggl \lbrace V_{h  a_1a_2,h  b_1b_2} 
          - 
                         V_{h  a_1a_2,h  b_2b_1} 
          \biggr \rbrace
   + 
   \sum_h \biggl \lbrace - V_{h  a_1a_2,b_1 h  b_2}
   \nonumber \\ &&  \qquad 
                         - V_{h  a_2a_1,b_2 h  b_1}
                         + V_{h  a_2a_1,b_1 h  b_2}
                         + V_{h  a_1a_2,b_2 h  b_1}
          \biggr \rbrace 
   \>.
\label{eq:VTNI_den}
\end{eqnarray}
The definition~(\ref{eq:VTNI_den}) has been inspired by the form of
second order (${\bf S}_2$) contributions to the binding-energy and has
the additional advantage of being fully anti-symmetric, so that all
procedures developed when dealing with the nucleon-nucleon
interaction~\cite{ref:2b-me}, can be naturally extended to handle the
three-nucleon interaction.  Note that the first two terms in
Eq.~(\ref{eq:VTNI_den}) are equivalent to the standard
density-dependent reduction of the three-body force.

We shall now detail the changes necessary to take into account the
effects of the three-nucleon interaction in the calculation of the
binding-energy and mean field using the coupled-cluster formalism.

%
%
\subsection{Binding Energy Corrections.}

We are only interested in the total binding energy when the wave
function satisfies the Hartree-Fock conditions.  Thus, it suffices to
compute
\begin{equation}
   \langle E \rangle 
   \ = \
   \langle 0 | {\bf T} | 0 \rangle 
   +
   \langle 0 | {\bf V} | 0 \rangle
   +
   \langle 0 | {\bf S}_2 {\bf V} | 0 \rangle
   +
   \langle 0 | {\bf S}_3 {\bf V}_{tni} | 0 \rangle
   \>.
\end{equation}
The first order corrections to the binding energy are due to the
expectation value of the three-nucleon interaction in the uncorrelated
ground state.  We have
\begin{eqnarray}
   \langle 0 | {\bf V}_{tni} | 0 \rangle
   & = & 
   \frac{1}{6} 
   \sum_{h_1h_2h_3} 
   \biggl \{
      V_{h_1h_2h_3,h_1h_2h_3}  
      - V_{h_1h_2h_3,h_1h_3h_2} 
   \nonumber \\ && 
      + V_{h_1h_2h_3,h_3h_1h_2}
      - V_{h_1h_2h_3,h_3h_2h_1} 
   \nonumber \\ && 
      + V_{h_1h_2h_3,h_2h_3h_1} 
      - V_{h_1h_2h_3,h_2h_1h_3}
   \biggr \}
   \>.
\label{eq:vtni_av_1}
\end{eqnarray}
Using the symmetries~(\ref{eq:vtni_symm_a}, \ref{eq:vtni_symm_b}), 
the last equation becomes
\begin{eqnarray}
   \langle 0 | {\bf V}_{tni} | 0 \rangle
   & = &
   \sum_{h_1h_2h_3}
   \biggl \{
      \frac{1}{6} \, V_{h_1h_2h_3,h_1h_2h_3} 
   \nonumber \\ &&      
      \, - \,
      \frac{1}{2} \, V_{h_1h_2h_3,h_1h_3h_2} 
      \, + \,
      \frac{1}{3} \, V_{h_1h_2h_3,h_3h_1h_2}
   \biggr \}
   \>.
   \nonumber \\
\label{eq:vtni_av}
\end{eqnarray}
We notice that only the first two terms in Eq.~(\ref{eq:vtni_av}) have
the form~(\ref{eq:vtni_calc}).  However, in this particular case, it
is a simple endeavor to calculate the missing third matrix element.
We find that the magnitude of this term is small compared to the sum
of the terms~(\ref{eq:vtni_calc}).

Second order contributions are calculated {\em exactly} as
\begin{equation}
   \langle 0 | {\bf S}_2 {\bf V} | 0 \rangle
   \ = \
   \frac{1}{4} \, 
   \sum_{p_1h_1 p_2h_2}  
       S_{p_1h_1, \, p_2h_2} 
       V^{tni, den}_{p_2h_2, \, p_1h_1}
   \>.
\end{equation}
At the present time, the third order corrections have not been
evaluated. Their magnitude is expected to be small, and their 
inclusion is definitely next-order in terms of the present expansion.

%
%
\subsection{Mean Field Corrections.}

In our approach to the coupled cluster formalism, the single particle
orbits are eigenfunctions of an mean-field Hamiltonian, defined as the
sum of a one-body kinetic energy term and a one-body mean-field
potential.  The later is not unique, and in the maximum overlap
hypothesis ${\bf S}_1=0$, the mean-field is defined as
\begin{equation}
   \langle 0 | {\bf U}_0 | 1p1h \rangle
   \ = \
   \langle 0 |
       \bigg \lbrace
             {\bf V}
             \, + \, \big[ {\bf S}_2, {\bf V} \big]
             \, + \, \big[ {\bf S}_3, {\bf V} \big]
       \bigg \rbrace
   | 1p1h \rangle
   \> .
\label{eq:mf}
\end{equation}
Correspondingly, the contributions due to the three-nucleon interaction
can be written as the sum of three terms
\begin{eqnarray}
   &&
   \langle 0 | {\bf V}_{tni} {\bf a}^{\dag}_p {\bf a}_h
             | 0 \rangle
   \nonumber \\ &&
   + \ 
   \langle 0 | \bigl [ {\bf S}_2, {\bf V}_{tni} \bigr ] 
                     {\bf a}^{\dag}_p {\bf a}_h
             | 0 \rangle
   \ + \
   \langle 0 | \bigl [ {\bf S}_3, {\bf V}_{tni} \bigr ] 
                     {\bf a}^{\dag}_p {\bf a}_h
             | 0 \rangle
   \> .
\label{eq:vtni_mf_0}
\end{eqnarray}
In leading order, 
the three-nucleon interaction correction of the mean-field is given as
\begin{eqnarray}
   \langle 0 | {\bf V}_{tni} \, {\bf a}^{\dag}_p {\bf a}_h
             | 0 \rangle
   & = & 
   \frac{1}{2} \,
   \sum_{h_1h_2} \, 
   \biggl \{
       V_{h_1h_2 \, h;h_1h_2 \,p} 
       - 
       V_{h_2 \, h \, h_1,h_2h_1 \, p} 
   \nonumber \\ &&
       +
       V_{h_1h_2,h;h_2,p,h_1}
       - 
       V_{h_1h_2 \, h,h_1 \, p \, h_2} 
   \nonumber \\ &&
       +
       V_{h_2 \, h \, h_1,h_1h_2 \, p} 
       - 
       V_{h_1h_2 \, h,h_2h_1 \, p}
   \biggr \}
   \> .
   \nonumber \\
\label{eq:vtni_mf_a1}
\end{eqnarray}
Again, using the symmetries~(\ref{eq:vtni_symm_a}, \ref{eq:vtni_symm_b}), 
Eq.~(\ref{eq:vtni_mf_a1}) becomes
\begin{eqnarray}
   \langle 0 | {\bf V}_{tni} \, {\bf a}^{\dag}_p {\bf a}_h
             | 0 \rangle
   & = &
   \sum_{h_1h_2} 
   \biggl \{
       \frac{1}{2} \, V_{h_1h_2 \, h,h_1h_2 \,p} 
       - 
       V_{h_1 h \, h_2,h_1h_2 \, p} 
   \nonumber \\ &&       
       +
       V_{h_1h_2 \, h,h_2 \, p \, h_1} 
       - 
       \frac{1}{2} \, V_{h_1h_2 \, h,h_2h_1 \, p}
   \biggr \}
   \> .
   \nonumber \\
\label{eq:vtni_mf_a}
\end{eqnarray}

The second- and third-order contributions in 
Eq.~(\ref{eq:vtni_mf_0}) look very similar when one uses the 
proposed reduction of the three-body force, Eq.~(\ref{eq:VTNI_den}), 
for the account of the \nph{3} correlations.
Then, by making use of the full anti-symmetry of ${\bf S}_2$ 
and ${\bf V}_{tni, den}$ we can show that the required corrections 
can be written as
\begin{eqnarray}
   &&
   {\displaystyle 
   \sum_{p_1p_2h_1h_2} } 
   \Biggl \{ 
       \biggl ( V_{h  p_1p_2,p  h_1h_2}
               - 2 \, V_{h  p_1p_2,h_1  p  h_2}
       \biggr ) 
       Z_{p_1p_2,h_1h_2}
   \nonumber \\ && 
   -
       \biggl ( V_{h  h_1h_2,p  p_1p_2}
              - 2 \, V_{h  h_1h_2,p_1  p  p_2}
       \biggr )  
       \frac{ \langle p_1 \bar h_1 | V_{2N} | h_2 \bar p_2 \rangle}
            {\epsilon_{ph}+\epsilon_{p_1h_1}+\epsilon_{p_2h_2}}
   \Biggr \}
   \> .
   \nonumber \\
\label{eq:vtni_mf_b}
\end{eqnarray}
To maintain the symmetry we have also added similar terms
to the $hh$- and $pp$-terms of the single particle Hamiltonian.

%
%
\section{Observables}
\label{sec:obs}

The results we report here are carried out in a $50\hbar\omega$
configuration space.  In general, the expectation value of any
arbitrary operator $A$ in the correlated ground-state $| \tilde 0
\rangle$ is obtained as
\[
   \bar{a} =
   \langle 0 \, | \,
             e^{\bf S} \, A \, e^{- \bf S} \, \tilde{\bf S}^{\dag}
             \, | \, 0 \rangle
   \>,
\]
where similarly to the cluster correlation operator ${\bf S}^{\dag}$,
the new operator $\tilde{\bf S}^{\dag}$ is expanded out in terms of
{\em ph}-creation operators
\[
   \tilde{\bf S}^{\dag} =
   \sum_n \frac{1}{n!} \tilde{S}_n {\bf O}^{\dag}_n
   \>.
\]
The amplitudes $\tilde{S}_n$ are calculated iteratively in terms of
the ${S}_n$ amplitudes~\cite{ref:paper_one}.

\begin{figure}
   \epsfxsize = 3.0in
   \centerline{\epsfbox{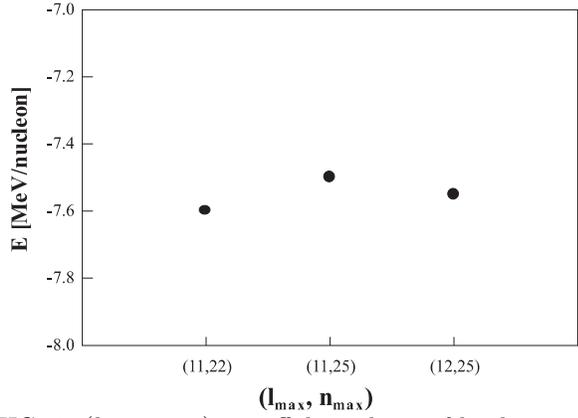}}
   \caption{($l_{max}$,$n_{max}$) cut-off dependence of binding energy 
            (using $v$18 and UIX).}
\label{fig:vtni_conv}
\end{figure}

The resulting binding energy and charge
radii for the Argonne $v$18 with/without the Urbana IX potential, are
shown in Table~\ref{table:energy_rms}. Note that the charge radii
correspond to the ``theoretical'' charge density as explained below.
The calculation is shown to be stable with respect to the two cut-off
parameters, $l_{max}$ and $n_{max}$, which control the size of the
configuration space as $N_{max}=2n_{max}+l_{max}$.
Figure~\ref{fig:vtni_conv} shows the binding energy for $^{16}$O
calculated using the $v$18 and Urbana IX interactions, as a function
of the ($l_{max}$, $n_{max}$) cut-off pair, and show the calculation
to be reasonably converged for $l_{max}=11,12$ and $n_{max}=22,25$,
same as the $v$18 based calculation of~\cite{ref:paper_one}.

%
%
\subsection{Charge Form Factor}

The charge form factor is calculated as the sum of a one- and two-body
operator. The one-body component is evaluated in the nonrelativistic
one-body Born-approximation picture, where the charge form factor is
given as
\[
   F_{L}(q) \ = \
   \langle
   \tilde 0 \, | \,
          \sum_k \, f_k(q^2) \ e^{i \vec{q} \cdot \vec r_k'}
          \, | \, \tilde 0 \rangle
   \>.
\]
Here $f_k(q^2)$ is the nucleon form factor, which takes into account
the finite size of the nucleon $k$.  In our calculation we use the
Iachello-Jackson-Lande~\cite{ref:ffn_Iach} nucleon-form factors.  Note
however that at the rather small values of~$q$ involved in the present
discussion, $q<4$ fm$^{-1}$, the differences between the various
models of the nucleon form factor are expected to be
small~\cite{ref:ffn_disc}.

\begin{figure}[h]
   \epsfxsize = 3.0in
   \centerline{\epsfbox{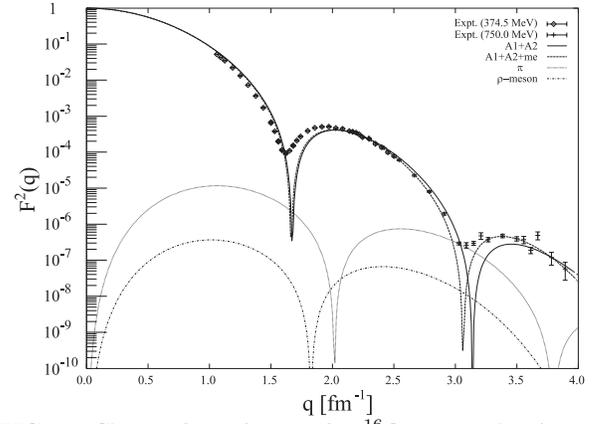}}
   \caption{Charge form factors for $^{16}$O,
            using the Argonne $v$18 and Urbana IX potentials.}
\label{fig:o16_ff}
\end{figure}

As seen from previous calculations~\cite{ref:Pieper_etal}, the main
contributions from the two-body charge density not related to
center-of-mass corrections is expected to come from the the $\pi$- and
$\rho$-exchange ``seagull" diagrams.  These are taken into account as
referred in the {\em model-independent} part of the ``Helsinki
meson-exchange model"~\cite{ref:ffn_disc}.  In this model the pion and
$\rho$-meson propagators are replaced by the Fourier transforms of the
isospin dependent spin-spin and tensor components of the $v$18 $NN$
interaction, in order to ensure that the exchange current operator
does satisfy the continuity equation together with the interaction
model. As seen from Fig.~\ref{fig:o16_ff}, the contributions of $\pi$-
and $\rho$-exchange give a measurable correction for $q>2$ fm$^{-1}$.
The order of magnitude of the $\pi$- and $\rho$-exchange contributions
compares well with the calculation of~\cite{ref:Pieper_etal} for
$^{16}$O using the Argonne $v$14 potential and Urbana VII
three-nucleon interaction.

Since the correlated ground state wave function $| \tilde 0 \rangle$
is not translationally invariant, extra care needs to be taken in
order to account for the effect of the center-of-mass motion on the
expectation value of the operator in the ground state.  Center-of-mass
corrections have been discussed in~\cite{ref:paper_two}, and rely on
the assumption that the motion of the intrinsic coordinates $\vec r_k'
= \vec r_k - \vec R_{cm}, k = 1 \ldots A$, and the center-of-mass are
not correlated, and that our correlated ground-state $| \tilde 0
\rangle$ provides indeed a good description of the internal structure
of the nucleus. 

Figure~\ref{fig:o16_ff} depicts the center-of-mass corrected charge
form factor both in the impulse approximation and including the
meson-exchange two-body component, together with the experimental
results for $q\le4$ fm$^{-1}$~\cite{ref:SickMcCarthy}.  These various
approximations of the theoretical charge form factor correspond to the
$v$18+UIX+$H_{CM}$ interaction in Table~\ref{table:energy_rms}.

\begin{figure}[h]
   \epsfxsize = 3.0in
   \centerline{\epsfbox{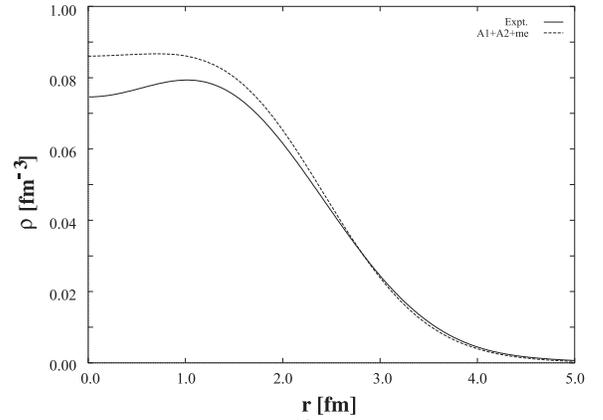}}
   \caption{Charge density for $^{16}$O,
            including meson exchange contributions.}
\label{fig:o16_den}
\end{figure}

Figure~\ref{fig:o16_den} shows the corresponding charge density
together with the ``experimental'' charge density as derived
in~\cite{ref:SickMcCarthy}. This calculation simply represents the
result of Fourier transforming the charge form-factor presented in
Fig.~\ref{fig:o16_ff}. However, in doing so we use predictions for a
momentum transfer greater than 4 fm$^{-1}$, where the present
calculation is not expected to offer reliable predictions.  This
results in uncertainties of the charge density distribution at short
distances $r$.

\begin{figure}[h]
   \epsfxsize = 3.0in
   \centerline{\epsfbox{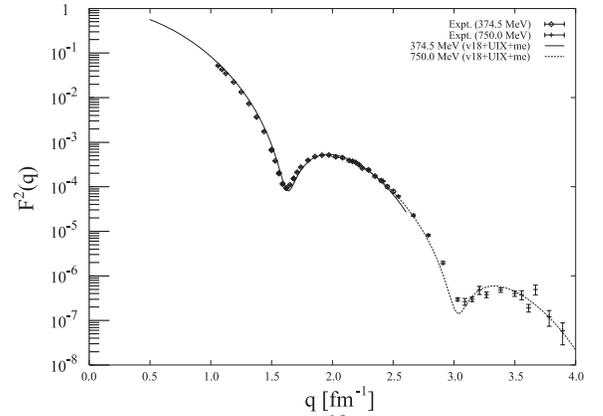}}
   \caption{Charge form factor for $^{16}$O,
            obtained in the DWBA picture ($v$18 + UIX + ME).}
\label{fig:o16_elastic}
\end{figure}

Also, the distortion effect related to the interaction of the electron
probe with the nuclear Coulomb field has not yet been accounted for.
Thus, we use the Distorted Wave Born Approximation (DWBA) to
calculated a new charge form factor~\cite{ref:DWBA}, which is depicted
in Fig.~\ref{fig:o16_elastic}. Note this calculation is now sensitive
to the energy of the incident electron as shown in
Fig.~\ref{fig:o16_elastic}.  This time the agreement is quite
impressive, and the calculation of the corresponding charge density
will give a result considerably closer to the ``experimental'' one. As
a result of this exercise, we conclude that it is better to compare
the results of the distorted charge form factor calculation with the
experimental charge form factor over the domain of momentum transfer
covered by the experiment.  In any event the charge form factor
represents the {\em primary} result of the experiment, and comparing
form factors is thus more reliable than comparing {\em
  model-dependent} charge densities.

Note that the persistent discrepancy at large $q$ may be a signature
of the break down of the nonrelativistic approach to calculating the
charge form factor. However, due to the various other approximations
made so far, a calculation based on the relativistic description of
the one-body charge form factor~\cite{ref:Sabine} may be too expensive
at this time.

%
%
\subsection{Coulomb Sum Rule}

The Coulomb sum rule, $S_L(q)$, represents the total integrated
strength of the longitudinal response function 
measured in inclusive electron scattering. The Coulomb sum rule is
related to the Fourier transform of the proton-proton distribution
function in the nuclear ground state~\cite{ref:SL_ppdf}. As such, this
quantity is sensitive to the short-range correlations induced by the
repulsive core of the $NN$ interaction.  In the nonrelativistic limit
we have~\cite{ref:Schiavilla_Carlson}
\begin{eqnarray}
   S_L(q)
   & = &
   \frac{1}{Z} \langle \tilde 0 | \rho^{\dag} \rho(q) | \tilde 0 \rangle
   -
   \frac{1}{Z} | \langle \tilde 0 | \rho(q) | \tilde 0 \rangle |^2
   \nonumber \\ 
   & \equiv &
   1 + \rho_{LL}(q) 
   -
   \frac{1}{Z} | \langle \tilde 0 | \rho(q) | \tilde 0 \rangle |^2
\label{eq:SL_eq}
   \>,
\end{eqnarray}
where $\rho(q)$ is the nuclear charge operator
\begin{equation}
   \rho(q)
   \ = \
   \frac{1}{2} \
   \sum_i^A \, e^{i \vec q \cdot \vec r_i} \ ( 1 + \tau_{z,\, i} )
   \>.
\end{equation}
In Eq.~(\ref{eq:SL_eq}), the longitudinal-longitudinal distribution
function~$\rho_{LL}(q)$ is given in terms of the proton-proton
two-body density as
\begin{equation}
   \rho_{LL}(q)
   \ = \
   \int d \vec r_1 \int d \vec r_2 \
   j_0(q |\vec r_1 - \vec r_2|) \ \rho^{(p,p)}(\vec r_1,\vec r_2)
   \>.
\end{equation}
We have outlined recently~\cite{ref:paper_one} the calculation of the
proton-proton two-body density in the coupled-cluster framework, as
the expectation value
\begin{eqnarray}
   &&
   \rho^{(p,p)}(\vec r_1,\vec r_2)
   \\ &&
   \ = \
   \frac{1}{4}
   \sum_{i,j}
   \langle \tilde 0 |
               \delta(\vec r_1 - \vec r_i) \delta(\vec r_2 - \vec r_j)
               ( 1 + \tau_{z,\, i} ) ( 1 + \tau_{z,\, j} )
   | \tilde 0 \rangle
   \>,
   \nonumber 
\end{eqnarray}
with the normalization
\begin{equation}
   \int d \vec r_1 \int d \vec r_2 \
   \rho^{(p,p)}(\vec r_1,\vec r_2) \ = \ Z - 1
   \>.
\end{equation}

\begin{figure}[h]
   \epsfxsize = 3.0in
   \centerline{\epsfbox{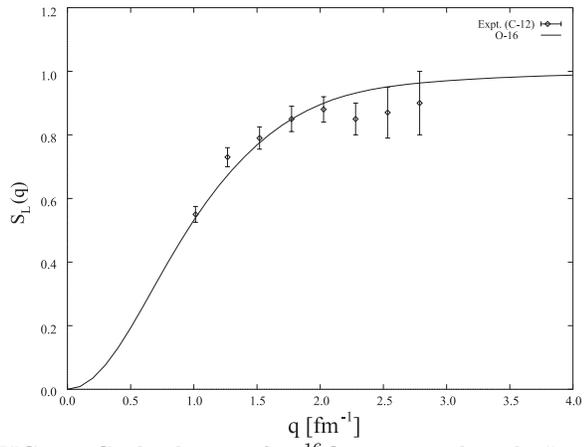}}
   \caption{Coulomb sum for $^{16}$O,
            compared with ``experimental" $^{12}$C data which include
            theoretically determined high-energy tail corrections.}
\label{fig:o16_sl}
\end{figure}

Figure~\ref{fig:o16_sl} depicts the results of the present calculation
for $^{16}$O. Agai, the theoretical calculation corresponds to the
$v$18+UIX+$H_{CM}$ interaction in Table~\ref{table:energy_rms}.  Since
no experimental data are available for $^{16}$O we compare these
results with the $^{12}$C experimental data of~\cite{ref:SL_c12},
where an estimate~\cite{ref:SL_tail} for contributions from large
$\omega$ has been added.  Preliminary theoretical results for $^{12}$C
obtained using the coupled-cluster method have been recently reported
in~\cite{ref:letter_O16} and shown to be close to the $^{16}$O result.
The large error bars on the experimental data are largely due to
systematic uncertainties associated with tail
contribution~\cite{ref:SL_2body}.

%
%
\section{Conclusions and Outlook}
\label{sec:end}

The goal of our effort is to make a contribution to the ongoing effort
of building realistic models of nuclear structure that explicitly
account for realistic correlations.  In a first stage we focus our
attention on obtaining a realistic description for the ground state of
a doubly-magic nucleus using the coupled-cluster method. In this
sense, the present calculation represents the most detailed
calculation available today, using the coupled-cluster method for a
nuclear system with $A > 8$. We base our calculation on what is
considered the best realistic description of the two- and three-body
interactions available today~\cite{ref:Pudliner_etal}.

At this stage we have proved conclusively that it is possible to
choose a large enough configuration space to handle the relatively
hard-core of the nucleon-nucleon interaction.  We believe that we have
also identified the dominant contributions of the three body
interaction, and we have introduced an effective two-body interaction,
accordingly.

An interesting conclusion from the charge form factor calculation has
to do with the importance of the distortion effect due to the
interaction between the electron and the Coulomb field of the nucleus.
In this context we have shown that it is desirable to compare the
distorted charge form factor with the experimental one, and to
de-emphasize the comparisons regarding the nuclear charge density,
which is not very surprising as the charge form factor represents the
primary result of the experiment.

In this paper we have confined ourselves to presenting a purely
nonrelativistic description of the charge form factor in~$^{16}$O.  In
the future it will be interesting to get a better understanding of the
importance of relativistic effects on the charge form factor
calculation. In this sense we plan to redo this calculation using the
relativistic description of the one-body charge form factor
of~\cite{ref:Sabine}, together with a consistent relativistic
derivation of the two-body meson-exchange density~\cite{ref:Amaro}.

With the calculation of the $^{16}$O ground state completed we intend
to extend our formulation to address the calculation of discrete
excited states as well as neighboring odd-even nuclei.  This is a
necessary step in the quest of modeling the (e,e'N) reaction, where
the final state has the asymptotic form of a distorted wave times a
discrete state of the (A-1) nucleus.  However, this is not a solution
to the Hamiltonian close to the origin, and thus the wave function
needs to be modified in the region of the origin.

%
%
\section*{acknowledgements}

This work was supported in part by the U.S. Department of Energy
(DE-FG02-87ER-40371).  The work of B.M. was also supported in part by
the U.S. Department of Energy under contract number DE-FG05-87ER40361
(Joint Institute for Heavy Ion Research), and DE-AC05-96OR22464 with
Lockheed Martin Energy Research Corp.  (Oak Ridge National
Laboratory).  Special thanks go to our colleagues F.~W.~Hersman,
M.~Holtrop, and their student T.~Streeter, J.~Kirkpatrick, and
D.~Protopopescu from the University of New Hampshire for setting up
the computer farm of 100 Intel Celeron processors at 500 MHz. This
allowed us to run several test cases simultaneously.  The authors
gratefully acknowledge useful conversations with John Dawson and David
Dean.

%
%

\appendix

%
%

\newpage

\begin{table}
   \caption{Energy expectation values and theoretical charge radii for 
            $^{16}$O.}
   \begin{center}
   \begin{tabular}{lcc}
      Interaction & B.E.  & r.m.s  \\
                  & [MeV/nucleon] & [fm$^{-1}]$ \\
      \hline
      $v$18       & -5.9  & 2.85       \\
      $v$18 + UIX & -7.7  & 2.74       \\
      $v$18 + UIX + $H_{CM}$
                  & -8.0  & 2.62       \\
      \hline
      expt.       & 8.00  & 2.73 $\pm$ 0.025
   \end{tabular}
   \end{center}
\label{table:energy_rms}
\end{table}

%
%
%

\end{document}